\documentclass{article}

\usepackage{arxiv}

\usepackage[utf8]{inputenc} 
\usepackage[T1]{fontenc}    
\usepackage{hyperref}       
\usepackage{url}            
\usepackage{booktabs}       
\usepackage{amsfonts}       
\usepackage{nicefrac}       
\usepackage{microtype}      
\usepackage{lipsum}		
\usepackage{graphicx}
\usepackage{natbib}
\usepackage{doi}
\usepackage{subeqnarray}
\usepackage{amsmath}
\usepackage{amstext}

\DeclareMathAlphabet\mathsfbi            {OT1}{cmss}{m}{sl}

\title{Analytic solution for pulse wave propagation in flexible tubes with application to patient-specific arterial tree}

\author{ \href{https://orcid.org/0009-0006-2765-2764}{\includegraphics[scale=0.06]{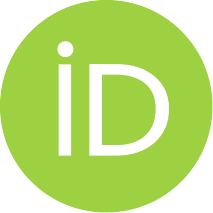}\hspace{1mm}Peishuo Wu} \\
	Department of Mechanics and Engineering Science \\
        \small{State Key Laboratory for Turbulence and Complex Systems}\\
	Peking University\\
	Beijing 100871, China \\
	\texttt{peishuo@stu.pku.edu.cn} \\
	\And
	\href{https://orcid.org/0000-0002-1099-8893}{\includegraphics[scale=0.06]{orcid.pdf}\hspace{1mm}Chi Zhu} \\
	Department of Mechanics and Engineering Science \\
        \small{State Key Laboratory for Turbulence and Complex Systems}\\
	Peking University\\
	Beijing 100871, China \\
        Nanchang Innovation Institute \\
        Peking University \\
        Nanchang 330008, China \\
	\texttt{chi.zhu@pku.edu.cn} \\
}

\renewcommand{\shorttitle}{\textit{Analytic solution for pulse wave propagation}}

\hypersetup{
pdftitle={A template for the arxiv style},
pdfsubject={q-bio.NC, q-bio.QM},
pdfauthor={David S.~Hippocampus, Elias D.~Striatum},
pdfkeywords={First keyword, Second keyword, More},
}

\begin{document}
\maketitle

\begin{abstract}
In this paper, we present an analytic solution for pulse wave propagation in a flexible arterial model with tapering, physiological boundary conditions and variable wall properties (wall elasticity and thickness). The change of wall properties follows a profile that is proportional to $r^\alpha$, where $r$ represents the lumen radius and $\alpha$ is a material coefficient. The cross-sectionally averaged velocity and pressure are obtained by solving a hyperbolic system derived from the mass and momentum conservations, and they are expressed in Bessel functions of order $(4-\alpha)/(3-\alpha)$ and $1/(3-\alpha)$, respectively. The solution is successfully validated by comparing it with numerical results from 3D fluid-structure interaction simulations. Subsequently, the solution is employed to study pulse wave propagation in an arterial model, revealing that the wall properties and the physiological outlet boundary conditions, such as the RCR model, play a crucial role in characterizing the input impedance and reflection coefficient. At low-frequency range, the input impedance is found to be insensitive to the wall properties and is primarily determined by the RCR parameters. At high-frequency range, the input impedance oscillates around the local characteristic impedance, and the oscillation amplitude varies non-monotonically with $\alpha$. Expressions for the input impedance at both low-frequency and high-frequency limits are presented. This analytic solution is also successfully applied to model flow inside a patient-specific arterial tree, with the maximum relative errors in pressure and flow rate never exceeding $1.6\%$ and $9.0\%$ when compared to results from 3D numerical simulation.
\end{abstract}


\section{Introduction} \label{sec:intro}
The pumping action of the heart creates pulse waves within the arterial system. The characteristics of pulse wave propagation is critical in understanding the behavior and functionality of arteries and, therefore, cardiovascular fitness \citep{safar2003,vandevosse2011}. Pulse wave propagation has been studied extensively through experiments \citep{moens1878, segers2000,bessems2008}, theoretical analysis \citep{korteweg1878, womersley1955, womersley1957, papadakis2011} and numerical simulations \citep{alastruey2011,mynard2015,charlton2019,zimmermann2021}.

In theoretical studies, the artery is commonly modeled as a straight flexible tube with uniform thickness and elasticity \citep{womersley1955, womersley1957,atabek1966,lighthill2001,flores2016}. However, the radius and thickness of the artery usually decrease along the blood flow direction, while the wall elasticity increases. It is known that these factors can change the local impedance of the artery and affect pulse wave propagation \citep{myers2004, vlachopoulos2011}. Also, arteries typically terminate at a bifurcation or connect to a vascular bed, resulting in an intricate outlet impedance that is commonly neglected in theoretical analysis.

\begin{figure}
  \centerline{\includegraphics[width=0.5\textwidth]{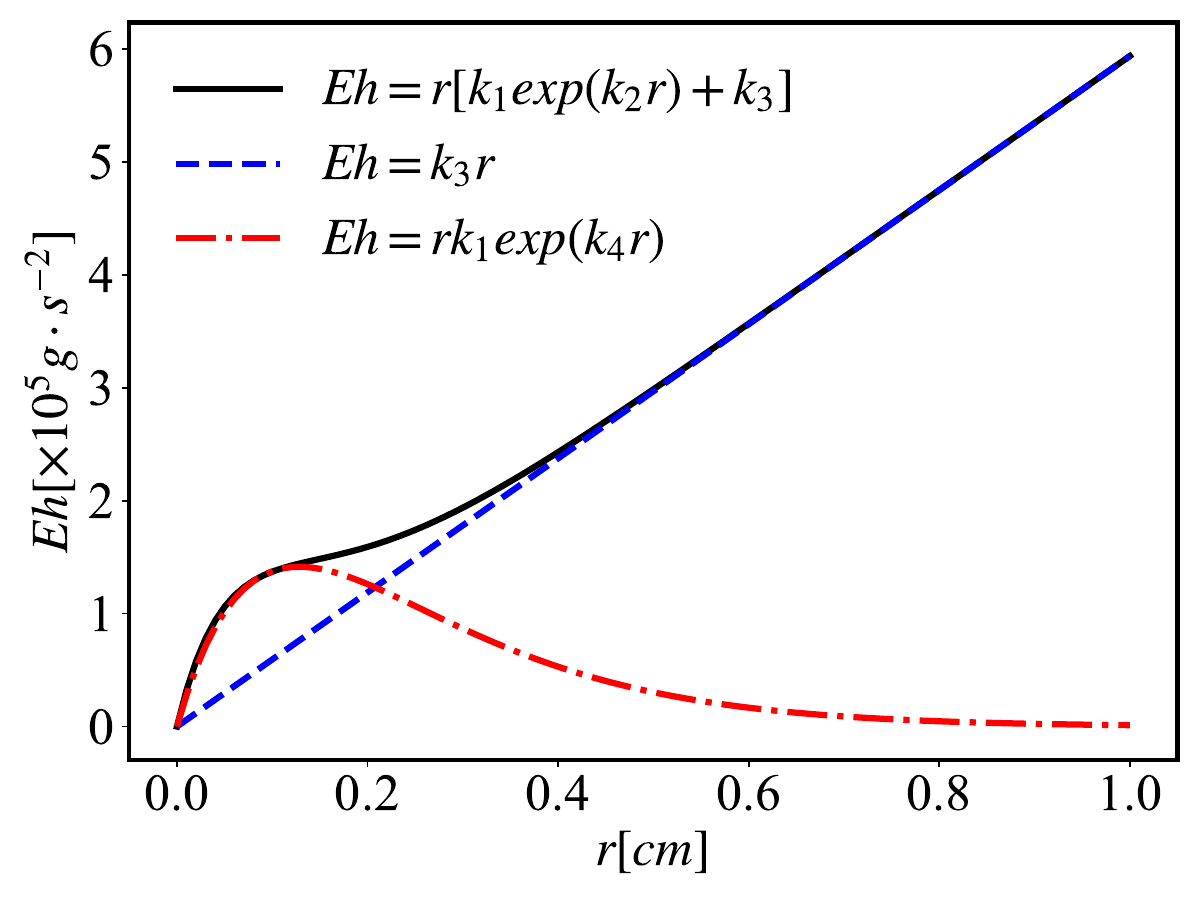}}
  \caption{Relationship between $E h$ and lumen radius of a healthy 30-year-old human subject with $k_{1}=$ $3 \times 10^{6} g \cdot cm^{-1} \cdot s^{-2} $, $k_{2}=-13.5 cm^{-1}$, $k_{3}=5.94 \times 10^{5} g \cdot cm^{-1} \cdot s^{-2}$ and $k_{4}=-7.8 cm^{-1}$ \citep{charlton2019}.}
\label{fig:Eh}
\end{figure}
For human arteries, tapering is usually mild, with the tapering angle not exceeding $1.5^\circ$ \citep{segers2000,papadakis2011}. In terms of wall properties, as will be shown in section \ref{sec:formula}, it is actually the product of the elastic modulus $E$ and the wall thickness $h$ that determines the overall property of the vessel wall. Hence, $Eh$ is sometimes called the arterial stiffness \citep{charlton2019}. Experimental research shows that $Eh$ can be approximated by the following function of the local lumen radius $r$
\begin{equation}
    Eh = r\left[ k_1 \exp (k_2r) + k_3 \right],
    \label{eq:Eh}
\end{equation}
where $k_1$, $k_2$ and $k_3$ are fitting parameters \citep{olufsen1999}. This wall property profile is widely adopted in 1D numerical studies \citep{mynard2015,charlton2019}. Figure \ref{fig:Eh} shows $Eh$ for a young healthy subject using this function. It can be seen that $E h$ can be approximated by a linear function when the vessel radius is greater than $2 mm$, which is true for most major arteries. It can be approximated by an exponential function instead when the vessel radius is less than $2 mm$. Moreover, other researchers \citep{reymond2009,willemet2015} have assumed that the relationship between $Eh$ and the local radius can be characterized by the following power-law function
\begin{equation}
    Eh = k_1 \overline{r}^{k_2}
\end{equation}
where $\overline{r}$ represents the time-averaged radius of the artery.

Among the aforementioned factors, tapering of the blood vessel probably receives the most attention. The first theoretical treatment of tapering is presented by \citet{evans1960}. The author found that it would cause constant reflection of the forward wave and claimed that only by considering tapering that we could explain the discrepancy between the pulse wave velocity predicted from existing theory and experimental measurements. However, in order to get an analytic solution, the vessel distensibility was assumed to be constant along the tapered vessel in this study, which is considered invalid from a modern perspective. \citet{patel1963} investigated the effect of tapering on pulse wave propagation in an animal experiment. By measuring the pressure-radius ($P-R$) relationship along the aorta of 30 dogs, they found that $\Delta P/\Delta R$, a measurement of local impedance, decreased as the mean radius reduced downstream of the aorta. \citet{lighthill1975} modeled the tapering of a vessel using a series of compact sections of straight tubes with stepwise diameter reduction and directly applied the solution from straight tubes to each section to study the pulse wave propagation. \citet{abdullateef2021} investigated the effect of tapering using 1D numerical simulations in time domain and confirmed that tapering would induce constant reflections which would led to increased pulse pressure amplification. They also studied the effect of modeling tapering with stepwise diameter reduction and concluded that this approach would cause artificial oscillations compared with smooth tapering. The wall properties (elasticity and thickness) were assumed to be uniform in their study. \citet{papadakis2011} started from the Navier-Stokes equation in the spherical coordinate system and derived the closed-form analytic solution for a tapered vessel with uniform wall properties. The pressure and velocity were expressed with the Bessel functions of orders 4/3 and 1/3. \citet{segers2000} conducted experiments with hydraulic models made up of tapering tubes with uniform wall properties and also carried out theoretical analysis using the transmission line theory. They concluded that the aortic wave reflection indices from in vivo measurements were resulted from the continuous wave reflection from tapering and local reflections from the branches.

As for wall properties, \citet{myers2004} accounted for both the geometric and elastic tapering in arteries by assuming the characteristic impedance and the propagation constant varied exponentially with the axial distance. The nonlinear Riccati equation for the input impedance was derived and solved to obtain the flow and pressure inside the model with the help of the transmission line method. \citet{wiens2021} recently studied the flow inside straight tubes with a tapered wall thickness using frequency domain analysis. They gradually varied the wall thickness along the axial direction while keeping the lumen radius and the elasticity unchanged, resulting in a varying wave velocity along the tube. They demonstrated that the change in wall thickness alone could  induce strong changes in the impedance and the wave propagation due to the change in wall compliance.

Another very important factor in the investigation of pulse waves is the proper treatment of the outlet boundary. Many theoretical studies of single tube models adopted the non-reflecting boundary condition \citep{womersley1955, papadakis2011}. On the other hand, an artery usually ends with branching or a vascular bed. \citet{taylor1966} studied the input impedance of the main artery connected to an artificial vascular bed and demonstrated that the vascular bed acted as a absorber to reduce the effect of reflections. Therefore, it is important to use proper outlet boundary conditions to incorporate the effect of downstream vessels so as to correctly capture pulse wave propagation in an arterial segment.

Studies that have taken tapering, variable wall properties and physiological boundary conditions into considerations are mainly 1D numerical studies \citep{bessems2008,reymond2009,xiao2014,willemet2015,mynard2015}. A theoretical analysis that includes all of these factors is still lacking. In this study, we present an analytic solution for the wave propagation in a flexible tapered arterial model with variable wall properties and physiological boundary conditions.

This paper is organized as follows. Section \ref{sec:formula} states the problem solved, including the governing equations, boundary conditions and wall properties. In section \ref{sec:analytic_solution}, we obtain the analytic solution in the velocity/pressure form using frequency domain analysis and validate it with results from 3D fluid-structure interaction (FSI) simulations. The analytic solution is subsequently used to analyze the characteristics of pulse wave propagation in section \ref{sec:analysis}, focusing on the impact of wall properties on impedance and reflection. The potentials of the obtained solution are demonstrated through its application to a patient-specific geometry in section \ref{sec:application}. Finally, section \ref{sec:conclusion} presents conclusions and discusses limitations of the current work.

\section{Problem statement} \label{sec:formula}

In this section, we define the problem solved by presenting the governing equations, boundary conditions and vessel wall properties for a canonical arterial model. Some key assumptions of the study are discussed.

\subsection{Governing equations}

\begin{figure}
  \centerline{\includegraphics[width=0.6\textwidth]{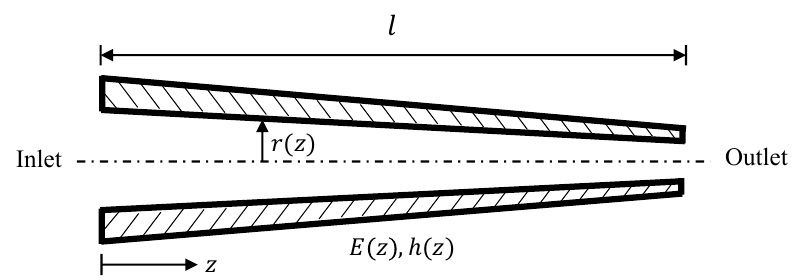}}
  \caption{Schematic of a tapered artery model with variable wall elasticity and thickness.}
\label{fig:geometry}
\end{figure}
We model the artery as a tapered axisymmetric tube of length $l$, as shown in figure \ref{fig:geometry}. The model has spatially distributed lumen (inner) radius $r(z)$, modulus of elasticity $E(z)$ and wall thickness $h(z)$, where $z$ is the axial coordinate. Following the work of \citet{papadakis2011}, we assume a linear tapering and the change of lumen radius along $z$ is given by
\begin{equation}
  r(z) = r_0 - bz,
  \label{eq:taper}
\end{equation}
where $r_0$ is the radius at the inlet and $b$ is a constant that characterizes the degree of tapering. For arteries, tapering is usually very mild and the maximum of $b$ is around 0.026 \citep{segers2000}. The blood is assumed to be incompressible. Starting from the mass and momentum conservations, the following classical 1D equations in the cylindrical coordinate system can be derived for the current model under the assumption that the axial displacement is negligible; there is no flow through the lumen wall along $z$ direction; and velocity and pressure are uniform in the cross section \citep{sherwin2003,vandevosse2011,figueroa2017}
\begin{subeqnarray}
    \frac{\partial A}{\partial t}+\frac{\partial(A v)}{\partial z}&=&0, \\ [3pt]
    \frac{\partial v}{\partial t}+v \frac{\partial v}{\partial z}+\frac{1}{\rho} \frac{\partial p}{\partial z}&=&\frac{f}{\rho A}.
    \label{eq:ns1}
\end{subeqnarray}
Here, $A$ is the area of the cross section; $f$ is the frictional force per unit length; $v$ and $p$ are the cross-sectionally averaged axial velocity and pressure, respectively.

To close this 1D model, we need an extra equation to describe the fluid-structure interaction between blood flow and vessel wall. This is achieved by adopting the tube-law \citep{sherwin2003,alastruey2011,papadakis2011}
\begin{equation}
    p=p_{ext}+\frac{4 E h}{3 r^{2}} u_r,
    \label{eq:tubelaw1}
\end{equation}
where $p_{ext}$ is the constant external pressure and $u_r$ is the radial displacement of the vessel wall. It is worth noting that equation \ref{eq:tubelaw1} is derived assuming small displacement ($u_r \ll r$) and the vessel wall to be incompressible, linear elastic, thin-walled and longitudinally tethered \citep{sherwin2003}. This set of equations \ref{eq:ns1}, \ref{eq:tubelaw1} have been widely used in numerical study of pulse wave propagation in arteries with tapering and variable wall properties \citep{sherwin2003,alastruey2011,vandevosse2011, figueroa2017}.

Equations \ref{eq:ns1} and \ref{eq:tubelaw1} form the governing equations in $(A,v,p)$ form. They are recast to $(v_r,v,p)$ form for easier manipulation in this study, which is achieved by noting that $\partial A /\partial t \approx 2\pi r v_r$ under the small displacement assumption.  $v_r$ is the radial velocity at the fluid-structure interface.
Following the findings from previous work \citep{sherwin2003,reymond2009}, the nonlinear term and viscous term have secondary contribution to the momentum conservation and thus are omitted from equation \ref{eq:ns1}b. To sum up, the governing equations utilized in this study are as follows
\begin{subeqnarray}
    r \frac{\partial v}{\partial z}+2 v_{r}+2 \frac{\partial r}{\partial z} v  &=& 0 \\ [3pt]
    \frac{\partial v}{\partial t}+\frac{1}{\rho} \frac{\partial p}{\partial z} &=& 0 \\ [3pt]
    \frac{\partial p}{\partial t} - \frac{4 E h}{3 r^{2}} v_{r} &=& 0
    \label{eq:ge}
\end{subeqnarray}

In this study, we focus on arteries with medium to large sizes. Based on the discussion in section \ref{sec:intro}, we assume that wall properties follow the general form $E h=\beta r^{\alpha}$ and limit the study to cases with $\alpha=0,~1~ \text{or}~ 2$ to facilitate discussion.

\subsection{Boundary conditions}

\begin{figure}
    \centerline{\includegraphics[width=0.5\textwidth]{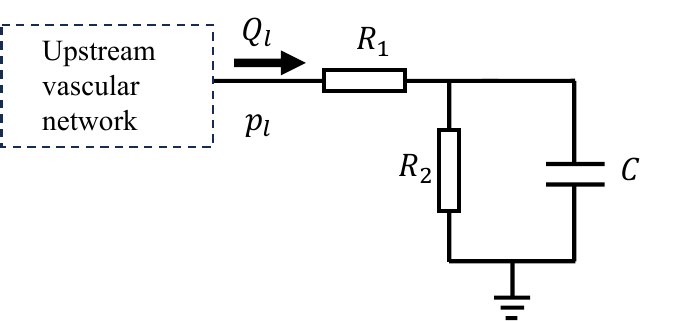}}
    \caption{The RCR boundary condition at the outlet. $C$ is the vascular compliance. $R_1$, $R_2$ are the proximal and distal resistance, respectively.}
  \label{fig:RCR}
\end{figure}

Proper boundary conditions are required to form a well-posed problem together with the governing equations. At the inlet, a commonly adopted boundary condition is a prescribed velocity profile $v_{in}(t)$ from in vivo measurements
\begin{equation}
    v_0(t)=v(0,t)=v_{in}(t).
    \label{eq:inletbc}
\end{equation}
Due to the pulsatile nature of the cardiovascular flow, $v_{in}(t)$ is usually periodic.

The outlet boundary condition represents the effect of downstream vascular networks on the current section, and it plays an important role in capturing the correct characteristics of pulse wave propagation. The downstream effect is usually modeled using lumped parameter models, and the RCR model (three-element Windkessel model, see figure \ref{fig:RCR}) is one of the most popular choices \citep{westerhof2009}. This model is composed of the vascular compliance $C$, the proximal resistance $R_1$ and the distal resistance $R_2$. These parameters can be tuned to match the physiological condition of a patient. The RCR boundary condition at the outlet ($x=l$) is governed by the following ordinary differential equation (ODE)
\begin{equation}
    \frac{\mathrm{d} p_{l}}{\mathrm{d} t} + \frac{p_{l}}{C R_{2}}=\frac{\left(R_{1}+R_{2}\right)}{C R_{2}} Q_{l} + R_{1} \frac{\mathrm{d} Q_{l}}{\mathrm{d} t},
    \label{eq:rcrbc}
\end{equation}
where flow rate $Q_{l}=\pi r_{l}^{2} v_{l}$, and $p_l$ and $v_l$ are average pressure and axial velocity at the outlet, respectively.

\section{Analytic solution} \label{sec:analytic_solution}

In this section, we present the closed-form solution to the problem. Solution for a single frequency mode is first derived, and then the time domain solution is obtained by the superposition of all frequency modes. Finally, the solution is validated with 3D numerical simulations.

\subsection{Solution for a single frequency mode}
\label{sec:sol_single_freq}

To solve the governing equations \ref{eq:ge}(a-c) with frequency domain analysis, we assume $p(z, t)=P(z) e^{\text{i} \omega t}$ and $ v(z, t)=V(z) e^{\text{i} \omega t}$, respectively. Replace $v_r$ in equation \ref{eq:ge}a with equation \ref{eq:ge}c and replace $Eh$ with $\beta r^{\alpha}$, we end up with the following equations in the frequency domain
\begin{subeqnarray}
    \text{i} \omega \frac{3 r^{2-\alpha}}{2 \beta} P+r \frac{\mathrm{d} V}{\mathrm{d} z}-2 b V=0, \\
    \text{i} \omega V+\frac{1}{\rho} \frac{\mathrm{d} P}{\mathrm{d} z}=0.
    \label{eq:ge_freq}
\end{subeqnarray}
Accordingly, the boundary conditions are transformed into the following form
\begin{subeqnarray}
    v(0,t)&=&V_{in} e^{\text{i} \omega t}\\
    p(l,t)&=&Z_{l} V_{l}e^{\text{i} \omega t}
\end{subeqnarray}
where
\begin{equation}
    Z_l=\frac{\left(R_{1}+R_{2}+\text{i} \omega R_{1} R_{2} C\right) \pi r_{l}^{2}}{1+\text{i} \omega C R_{2}}
    \label{eq:Zl}
\end{equation}
is the impedance of the RCR boundary.

Substitute the pressure in equation \ref{eq:ge_freq}b with equation \ref{eq:ge_freq}a and change the partial derivative of $z$ to that of $r$ following the linear tapering relation
\begin{equation}
    \frac{\mathrm{d}}{\mathrm{d} z}=-b \frac{\mathrm{d}}{\mathrm{d} r} \quad, \frac{\mathrm{d}^{2}}{\mathrm{d} z^{2}}=b^{2} \frac{\mathrm{d}^{2}}{\mathrm{d} r^{2}}.
\end{equation}
We obtain a second-order ODE of $V(r)$
\begin{equation}
    r^{2} \frac{\mathrm{d}^{2} V}{\mathrm{d} r^{2}} +(\alpha+1) r \frac{\mathrm{d} V}{\mathrm{d} r} + \left[ \frac{3 \rho \omega^{2}}{2 \beta b^{2}} r^{3-\alpha}-(4-2 \alpha)\right] V=0.
    \label{eq:besseleq}
\end{equation}
With the following transformation
\begin{equation}
    y = r^{\frac{\alpha}{2}}V, \quad \varepsilon = \sqrt{\frac{6\rho}{\beta}}\frac{\omega}{b(3-\alpha)}r^{\frac{3-\alpha}{2}}, \quad \nu = \frac{4-\alpha}{3-\alpha} , \nonumber
    \label{eq:transformation}
\end{equation}
this ODE can be rewritten into the standard Bessel equation of order $\nu$ in $y(\varepsilon)$
\begin{equation}
     \varepsilon ^{2}\frac{\mathrm{d}^{2}y}{\mathrm{d}\varepsilon^{2}}+\varepsilon \frac{\mathrm{d}y}{\mathrm{d}\varepsilon } +(\varepsilon ^{2}-\nu^{2})y=0.
\end{equation}
Therefore, equation \ref{eq:besseleq} has the following general solution \citep{bowman2012}
\begin{equation}
    V=r^{-\frac{\alpha}{2}}\left[c_{1} J_{\nu}(\varepsilon)+c_{2} Y_{\nu}(\varepsilon)\right]
    \label{eq:v_sol}
\end{equation}
where $J_{\nu}$ and $Y_{\nu}$ are Bessel functions of the first and second kind, and $c_{1}$ and $c_{2}$ are undetermined constants. Pressure can be easily obtained from equation \ref{eq:ge_freq}a
\begin{equation}
    P= -\text{i} \sqrt{\frac{2\rho\beta}{3r}} \left[c_{1} J_{\nu-1}(\varepsilon)+c_{2} Y_{\nu-1}(\varepsilon)\right].
    \label{eq:p_sol}
\end{equation}
Since $V$ and $P$ satisfy the RCR boundary condition at the outlet, we get $c_{1}=F c_{2}$, where
\begin{equation}
    F= - \frac{ \text{i} Z_{l} r_{l}^{\frac{1-\alpha}{2}} Y_{\nu}\left(\varepsilon_{l}\right)- B Y_{\nu-1}\left(\varepsilon_{l}\right)} { \text{i} Z_{l} r_{l}^{\frac{1-\alpha}{2}} J_{\nu}\left(\varepsilon_{l}\right) - B J_{\nu-1}\left(\varepsilon_{l}\right)}
    \label{eq:F}
\end{equation}
with $B=\sqrt{2 \rho \beta/3}$. Taking the inlet boundary condition into consideration, it is solved that
\begin{equation}
    c_{1}=V_{i n} r_{0}^{\frac{\alpha}{2}} \frac{F}{F J_{\nu}\left(\varepsilon_{0}\right)+Y_{\nu}\left(\varepsilon_{0}\right)}, \quad c_{2}=V_{i n} r_{0}^{\frac{\alpha}{2}} \frac{1}{F J_{\nu}\left(\varepsilon_{0}\right)+Y_{\nu}\left(\varepsilon_{0}\right)}.
\end{equation}

To sum up, the analytic solution for a single frequency mode is
\begin{subeqnarray}
    &V(z,\omega) = V_{i n} \left(\frac{r}{r_{0}}\right)^{-\frac{\alpha}{2}}  \frac{I_v(\varepsilon)}{I_v(\varepsilon_{0})}, \\
    &P(z,\omega) = -\text{i} B V_{in} \left(\frac{r}{r_{0}^\alpha}\right)^{-\frac{1}{2}}   \frac{I_p(\varepsilon)}{I_v(\varepsilon_{0})},
    \label{eq:solution}
\end{subeqnarray}
with $F$ being defined by equation \ref{eq:F} and
\begin{align}
    I_p(\varepsilon)=F J_{\nu-1}(\varepsilon)+Y_{\nu-1}(\varepsilon), \quad I_v(\varepsilon)=F J_{\nu}(\varepsilon)+Y_{\nu}(\varepsilon), \nonumber \\
    B=\sqrt{\frac{2 \rho \beta}{3}}, \quad \varepsilon = \sqrt{\frac{6\rho}{\beta}}\frac{\omega}{b(3-\alpha)}r^{\frac{3-\alpha}{2}}. \nonumber
\end{align}
Setting $\alpha=0$ in equation \ref{eq:solution} results in a solution that is similar to the one obtained by \citet{papadakis2011}, which is for a tapered vessel with uniform wall properties. They are all expressed in Bessel functions of order $4/3$ and $1/3$. But differences exist as they are derived under different coordinate systems and complicated boundary conditions are considered in the current study.

\subsection{Analytic solution in time domain}

Through the discussion in section \ref{sec:sol_single_freq}, it can be seen that the governing equations and the boundary conditions are all linear with regard to the primary variables. Therefore, the velocity and pressure solutions correspond to an arbitrary periodic inlet velocity profile can be obtained by the superposition of all frequency modes. An inlet velocity profile with period $T$ can be expanded into Fourier series
\begin{equation}
    v_{i n}(t)=\sum\limits_{n=-\infty}^{\infty} V_{n}^{in} e^{\text{i} \omega_n t}
\end{equation}
where
\begin{align}
    V_{n}^{in}=\frac{1}{T} \int_{0}^{T} v_{in}(t) e^{ -\text{i} \omega_n t} d t, \quad
    \omega_n=2 \pi n / T. \nonumber
\end{align}
The same operation can be carried out for the velocity and pressure solutions
\begin{equation}
    v(z,t)=\sum\limits_{n=-\infty}^{\infty} V_{n}(z) e^{\text{i} \omega_n t}, \quad
    p(z,t)=\sum\limits_{n=-\infty}^{\infty} P_{n}(z) e^{\text{i} \omega_n t}.
\end{equation}
For $n>0$, the expressions for $V_n(z)$ and $P_n(z)$ are provided by equation \ref{eq:solution}, while $n=0$ corresponds to the steady flow solution, which is governed by the following equations
\begin{subeqnarray}
    r \frac{\mathrm{d} V_{0}}{\mathrm{d} z}+2 \frac{\mathrm{d} r}{\mathrm{d} z} V_{0} &= 0, \\
    V_{0} \frac{\mathrm{d} V_{0}}{\mathrm{d} z}+\frac{1}{\rho} \frac{\mathrm{d} P_{0}}{\mathrm{d} z} &= 0.
\end{subeqnarray}
Note that the nonlinear term is included here, which we find to improve the accuracy of the pressure prediction. Combining with the boundary conditions, we can obtain the steady state solution as
\begin{subeqnarray}
    &V_{0}(z)=V_{0}^{in} \frac{r_{0}^{2}}{r^{2}}, \\
    &P_{0}(z)=V_{0}^{in} \pi r_{0}^{2} (R_1 + R_2)+\frac{1}{2} \rho\left(V_{0}^{in} \frac{r_{0}^{2}}{r_{l}^{2}}\right)^{2}-\frac{1}{2} \rho\left(V_{0}^{in} \frac{r_{0}^{2}}{r^{2}}\right)^{2}.
    \label{eq:steady_solution}
\end{subeqnarray}
Equation \ref{eq:steady_solution}a is a direct result of mass conservation, while equation \ref{eq:steady_solution}b is essentially the Bernoulli equation. The first term on the right hand side of equation \ref{eq:steady_solution}b represents the outlet pressure because RCR boundary is reduced to resistance boundary for steady flow, and the second term is the difference in kinetic energy .

Finally, the time domain solution is
\begin{subeqnarray}
    v(z,t) &=& V_{0}(z) + \mbox{Re}\left\{ \sum\limits_{n=1}^{\infty} 2 V_{n}^{in}\left(\frac{r}{r_{0}}\right)^{-\frac{\alpha}{2}} \frac{I_v(\varepsilon)}{I_v(\varepsilon_0)} e^{\text{i} \omega_n t} \right\}, \\
    p(z,t) &=& P_{0}(z) + \mbox{Re}\left\{ \sum\limits_{n=1}^{\infty} -\text{i}2 BV_{n}^{in} \left(\frac{r}{r_{0}^\alpha}\right)^{-\frac{1}{2}} \frac{I_p(\varepsilon)}{I_v(\varepsilon_0)} e^{\text{i} \omega_n t} \right\}.
    \label{eq:time_domain_solution}
\end{subeqnarray}
In all cases presented in this paper, we retain the first 20 terms of the series. It has been confirmed that any further increase in $n$ beyond 20 results in negligible improvement to the solution.

\subsection{Validation of the analytic solution}

\begin{table}
    \begin{center}
\def~{\hphantom{0}}
    \begin{tabular}{ c c }
    \multicolumn{2}{c}{Wall properties} \\
    $Eh(g\cdot s^{-2})$     & $\beta=5.94\times 10^5, \alpha=1$ \\
    Poisson's ratio            & 0.5 \\
    $\rho_s(g\cdot cm^{-3})$ & 1.0 \\
    \hline
    \multicolumn{2}{c}{Fluid properties} \\
    $\mu (g \cdot cm^{-1} \cdot s^{-1})$ & 0.04 \\
    $\rho (g\cdot cm^{-3})$ & 1.06 \\
    \hline
    \multicolumn{2}{c}{Vessel geometry} \\
    $r_{0}(cm)$ & 0.4 \\
    $r_{l}(cm)$ & 0.2 \\
    $l(cm)$ & 12.6 \\
    \hline
    \multicolumn{2}{c}{Boundary conditions} \\
    Inlet &  Prescribed flow rate with $T=1.1s$ \\
    Outlet & $R_1, R_2\left(g \cdot {cm}^{-4} \cdot s^{-1}\right): 6854.8,14330$ \\
     & $C\left(g^{-1} \cdot {cm}^{4} \cdot s^2\right): 1.7529\times 10^{-5}$ \\
    \end{tabular}
    \caption{Parameters of tapered carotid artery model.}
    \label{tab:para_carotid}
    \end{center}
\end{table}

Flow through a tapered carotid artery model with the same geometric configuration as figure \ref{fig:geometry} is used to validate the analytic solution. All relevant parameters are summarized in table \ref{tab:para_carotid}. The analytic solution is compared with a 3D FSI simulation using coupled momentum method (CMM) \citep{figueroa2006}, which is implemented in the open-source software svFSI \citep{zhu2022}. CMM has recently been rigorously verified with Womersley's deformable wall analytical solution \citep{filonova2020}. The mesh resolution and the time step size follow the same settings as in \citet{xiao2014}.

We quantify the differences between the analytic solution and the 3D simulation results using the following errors
\begin{align}
    &\eta_{P, a v g} = \frac{1}{N} \sum\limits_{i=1}^{N}\left|\frac{P_{i}^{a}-P_{i}^{c}}{P_{i}^{c}}\right|, &\quad
    &\eta_{Q, a v g} = \frac{1}{N} \sum\limits_{i=1}^{N}\left|\frac{Q_{i}^{a}-Q_{i}^{c}}{max _{j}\left(Q_{j}^{c}\right)}\right|, \nonumber \\
    &\eta_{P, max } = max _{i}\left|\frac{P_{i}^{a}-P_{i}^{c}}{P_{i}^{c}}\right|, &\quad
    &\eta_{Q, max } = max _{i}\left|\frac{Q_{i}^{a}-Q_{i}^{c}}{max _{j}\left(Q_{j}^{c}\right)}\right|. \nonumber
\end{align}
$N$ is the number of sampling time points and is set to 125 here. $ P_{i}^{a}$ and $Q_{i}^{a}$ are the pressure and flow rate calculated analytically, while $P_{i}^{c}$ and $Q_{i}^{c}$ are the mean pressure and flow rate on the cross section from CMM. $\eta_{a v g}$ reports the average relative error, while $\eta_{max }$ is the maximum relative error. In order to avoid dividing by small values in flow rate comparison, we divide the errors by the maximum flow rate for normalization \citep{xiao2014}.

\begin{figure}
    \centerline{\includegraphics[width=\textwidth]{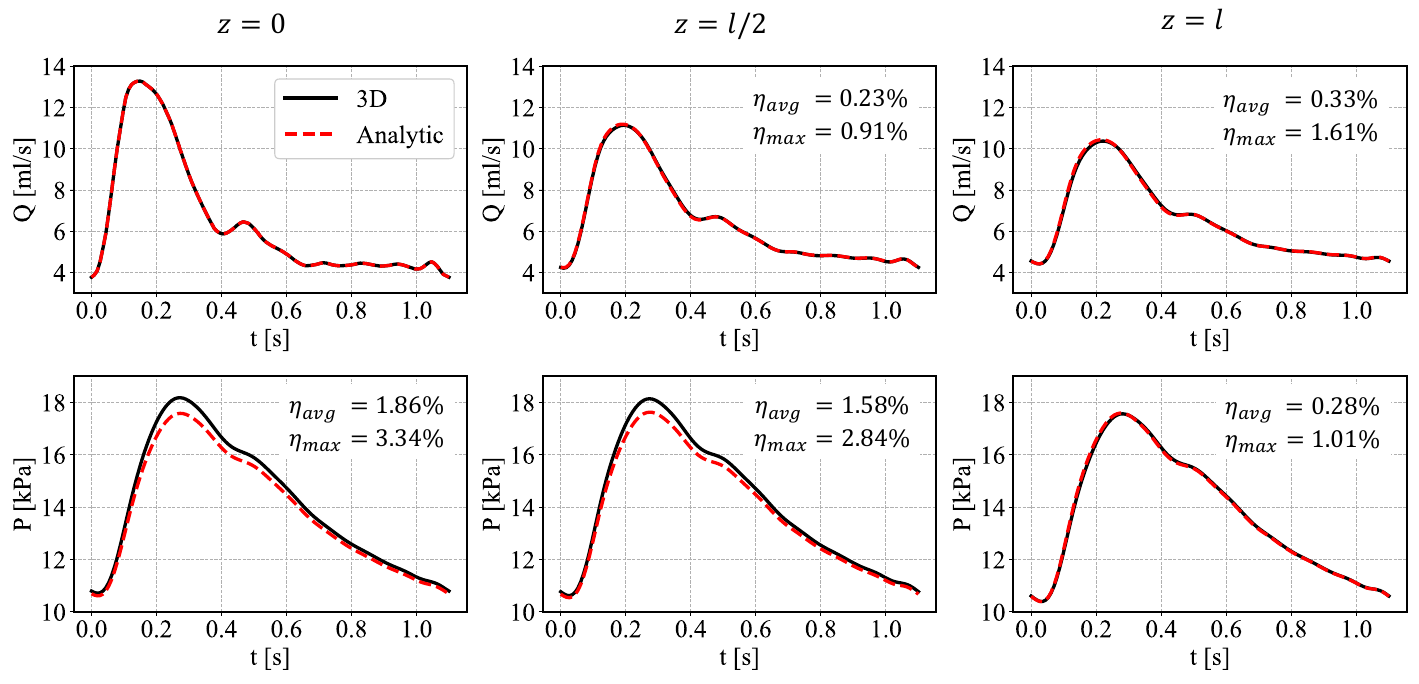}}
    \caption{Comparison between the analytic solutions and 3D simulation results for the tapered carotid with $\alpha=1$. Top row is the flow rate comparison, and bottom row is the pressure comparison.}
  \label{fig:carotid-1}
\end{figure}
The comparison of flow rate and pressure at the inlet, mid-section and outlet of the carotid artery model is summarized in figure \ref{fig:carotid-1}. Since the flow rates are prescribed at the inlet in both cases, they match exactly. Though errors in flow rate increase slowly towards the outlet, the values predicted by the analytic solution are still in excellent agreement with the 3D simulation results, with the maximum error being $1.61\%$ at the outlet. Moreover, the average relative errors of pressure never exceed $1.86 \%$, and the maximum relative errors remain under $3.4\%$. Contrary to flow rate, the errors of pressure decrease gradually from the inlet to the outlet. Since the RCR boundary condition is given at the outlet, the pressure is directly calculated from the flow rate there. Towards the inlet, the errors caused by omitting the fluid viscosity and nonlinear term likely accumulate to cause the slightly larger difference in pressure values.

\begin{figure}
    \centerline{\includegraphics[width=\textwidth]{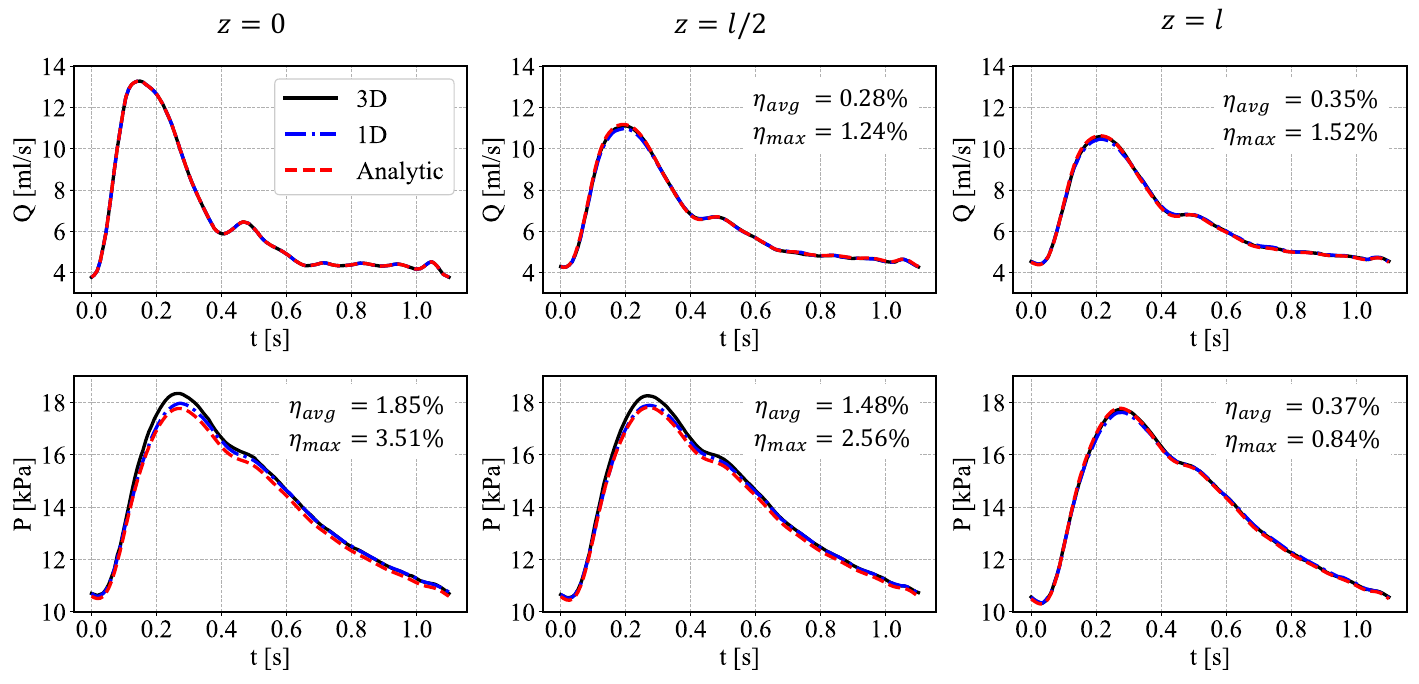}}
    \caption{Comparison between the analytic solutions and 1D, 3D simulation results for the tapered carotid with uniform wall properties. 1D results are extracted from \citet{xiao2014} and $\alpha=0$, $\beta=2.1\times 10^5$.}
  \label{fig:carotid-0}
\end{figure}
In addition to the 3D FSI simulation results, the analytic solutions are also compared with the results reported in \citet{xiao2014}, wherein they used 1D numerical simulations to study the pulse wave propagation in the same geometry but with uniform wall properties (see figure \ref{fig:carotid-0}). Overall, results predicted by the analytic solution are in great agreement with those from numerical simulations with either uniform or variable wall properties.

\section{Theoretical analysis of pulse wave propagation} \label{sec:analysis}

\begin{figure}
    \centerline{\includegraphics[width=0.9\textwidth]{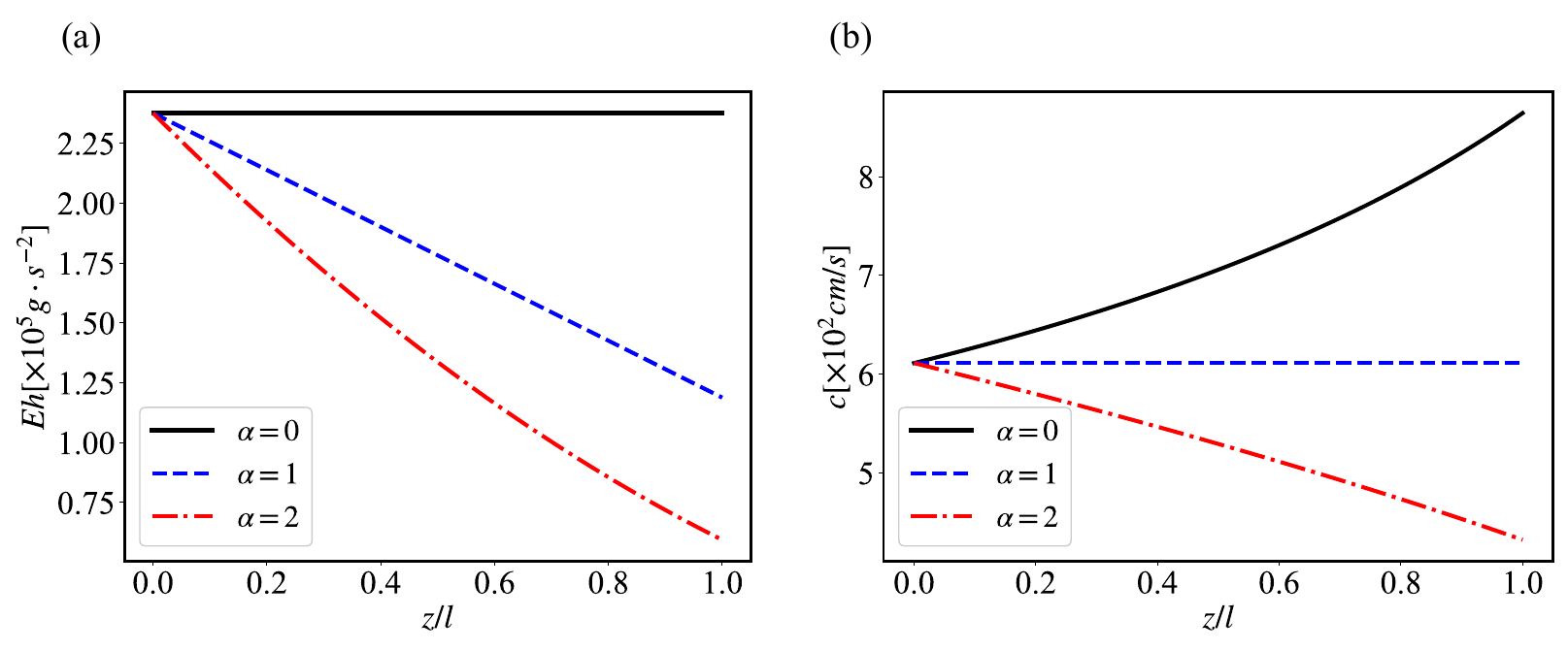}}
    \caption{Distribution of (a) wall properties and (b) wave velocity along the carotid artery model. Here, $\beta=2.376\times 10^5/r_0^\alpha$ so that all three cases have the same $Eh$ value at the inlet. Other parameters are from table \ref{tab:para_carotid}.}
  \label{fig:Eh-PWV-alpha}
\end{figure}

In this section, the analytic solution is used to analyze the effect of wall properties on pulse wave propagation in flexible tubes. As shown in figure \ref{fig:Eh-PWV-alpha}a, we focus on cases where $\alpha = 0,$ $1$ and $2$. $Eh$ value is kept the same at the inlet of all three cases.

\subsection{Wave propagation velocity}

The governing equation \ref{eq:ge} can be rewritten into the following form
\begin{equation}
    \frac{\partial \boldsymbol{U}}{\partial t}+\mathsfbi{A} \frac{\partial \boldsymbol{U}}{\partial z}=\mathsfbi{B} \boldsymbol{U}
\end{equation}
where
\begin{equation}
    \boldsymbol{U}=\left[\begin{array}{l}
    p \\
    v
    \end{array}\right], \quad \mathsfbi{A}=\left[\begin{array}{cc}
    0 & \frac{2 E h}{3 r} \\
    \frac{1}{\rho} & 0
    \end{array}\right], \quad \mathsfbi{B}=\left[\begin{array}{cc}
    0 & \frac{4 E h b}{3 r^{2}} \\
    0 & 0
    \end{array}\right].
\end{equation}
This apparently forms a system of hyperbolic equations, and the wave propagation velocity can be obtained by solving for the eigenvalues of matrix $\mathsfbi{A}$ \citep{papadakis2011,alastruey2012}, which is
\begin{equation}
    c= \pm \sqrt{\frac{2 E h}{3 \rho r}}= \pm \sqrt{\frac{2 \beta}{3 \rho}} r^{\frac{\alpha-1}{2}}.
    \label{eq:wave-velocity}
\end{equation}
It can be seen from figure \ref{fig:Eh-PWV-alpha} that as $\alpha$ increases, the $Eh$ value decreases at the same axial location of the model. On the other hand, wave velocity increases along the model when $\alpha=0$, while it decreases for $\alpha=2$. $\alpha=1$ is a special case where the wave velocity remains constant. It is worth noting that the wave velocity expressed in $Eh$ is consistent with the well-known Moens-Korteweg equation \citep{korteweg1878, moens1878, alastruey2012}, which is derived for straight tubes without tapering. \citet{papadakis2011} derived the wave velocity for a tapered vessel with uniform wall properties using spherical coordinates. The result included a correction term of second order $O(\theta^2)$ due to tapering, where $\theta=\arctan(b)$. In arterial systems, this correction term is negligible as the tapering angle is usually less than $1.5^\circ$.

\subsection{Input Impedance} \label{sec:impedance}

\begin{figure}
    \centerline{\includegraphics[width=0.9\textwidth]{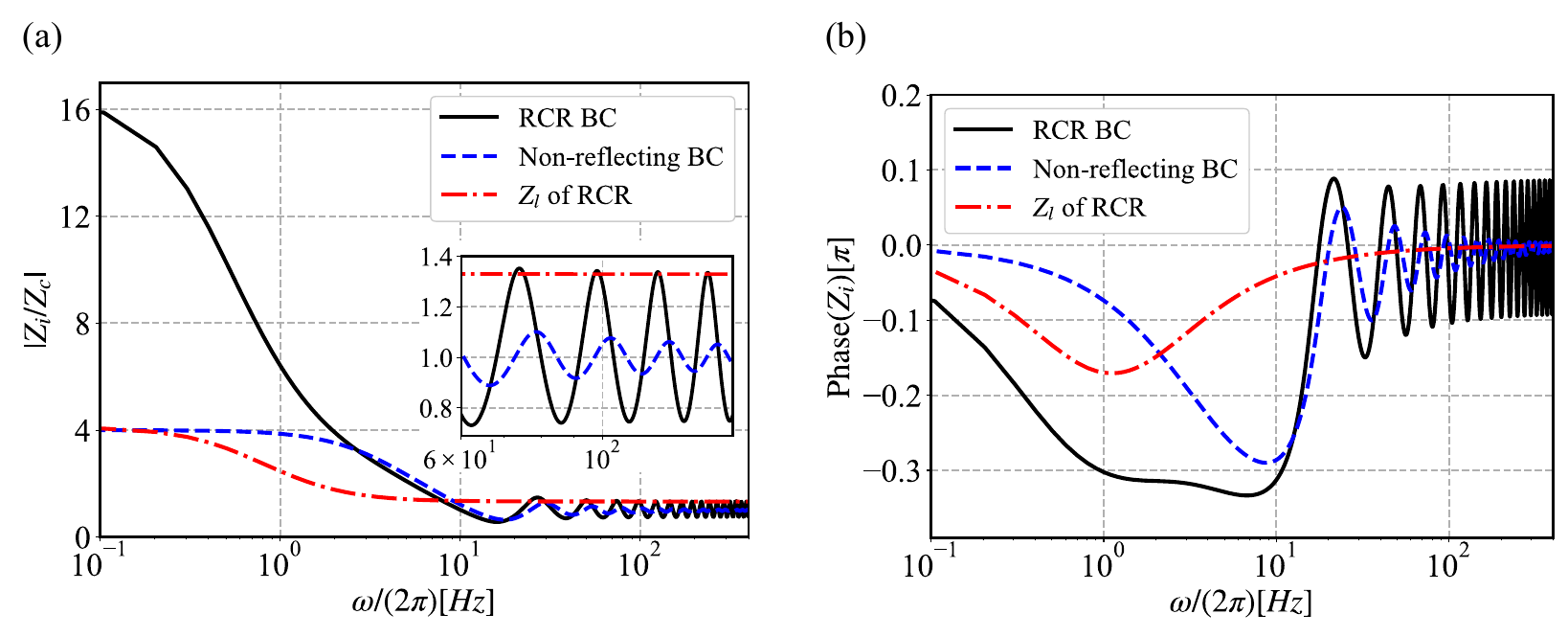}}
    \caption{(a) Magnitude and (b) phase of the input impedance of the tapered carotid artery model. Here, the magnitude is normalized by the characteristic impedance at the inlet. Parameters from table \ref{tab:para_carotid} are used.}
    \label{fig:Zi-alpha-1}
\end{figure}

With the wave velocity, the characteristic impedance can be expressed as $Z_c = \rho c$ \citep{westerhof2010}. It is a representation of the local wave transmission characteristic of the system without considering any reflections.

Moreover, based on the analytic solution \ref{eq:solution}, we have
\begin{equation}
    P = -\text{i} B r^{\frac{\alpha-1}{2}} \frac{I_p(\varepsilon)}{I_v(\varepsilon)} V = -\text{i} \rho c \frac{I_p(\varepsilon)}{I_v(\varepsilon)} V=Z V,
    \label{eq:input_imped}
\end{equation}
where $Z$ is the impedance \citep{westerhof2010}. $Z$ is a function of both axial location and frequency, while $Z_c$ is a function of axial location only. $Z$ evaluated at $z=0$ is referred to as the input impedance $Z_i$. Compared with the characteristic impedance, the additional coefficient $-\text{i}I_p/I_v$ in $Z_i$ measures the influence of the reflected waves caused by tapering, wall property change and outlet boundary condition. If $I_p/I_v=\text{i}$, the input impedance is equal to the characteristic impedance, i.e. there is no wave reflection from downstream of the inlet. This is discussed in detail below.

\begin{figure}
    \centerline{\includegraphics[width=0.9\textwidth]{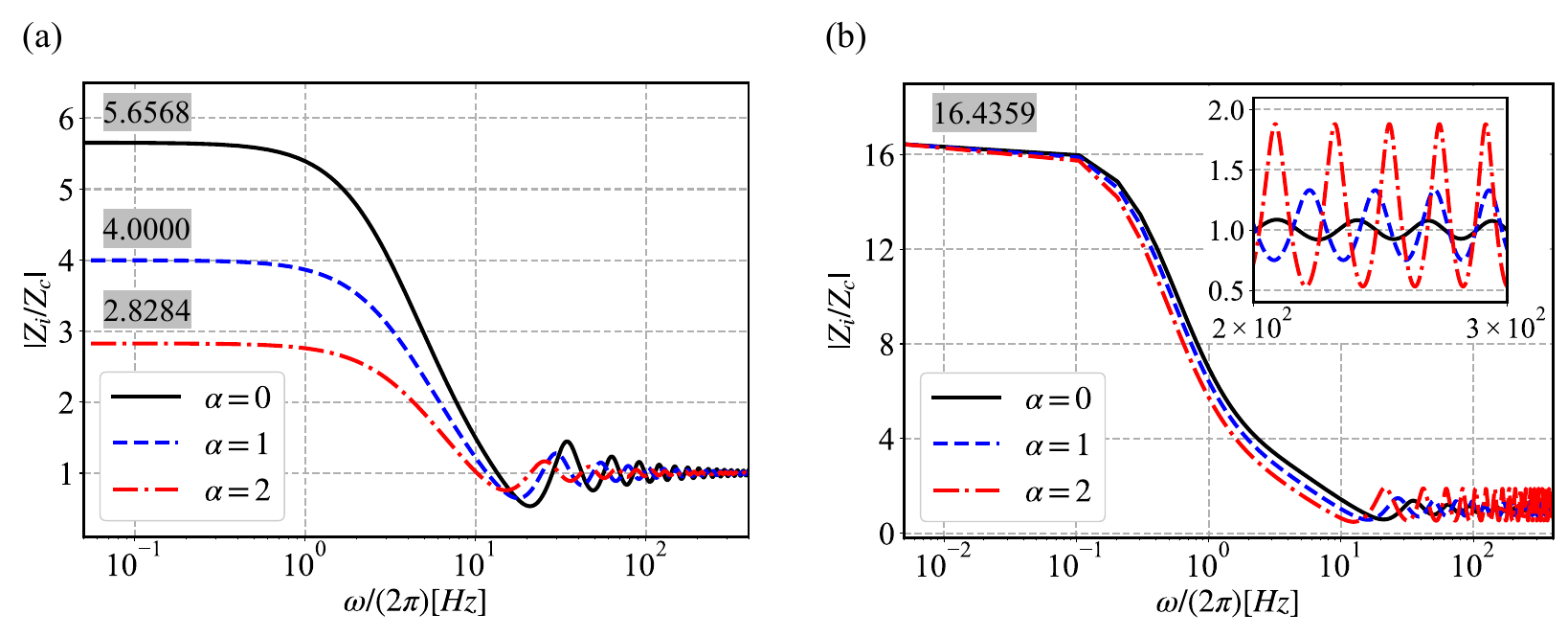}}
    \caption{Magnitude of the input impedance of the tapered carotid artery model with different $\alpha$.  (a) Non-reflecting boundary condition is employed; (b) RCR boundary condition is employed. The shaded numbers are the asymptotic values as $\omega \rightarrow 0$ predicted by equation \ref{eq:zero-limit}.}
    \label{fig:Zi-all}
\end{figure}

The change of input impedance with frequency is of great interest in cardiovascular research \citep{murgo1980,taylor1966}. Figure \ref{fig:Zi-alpha-1} plots the input impedance of the carotid artery model with $\alpha=1$. Two different boundary conditions are considered here
\begin{subeqnarray}
    Z_l&=&\rho c_l, \quad \quad\quad\quad\quad\quad\quad\quad\quad\quad  \mathrm{for \ non-reflecting \ boundary}; \\
    Z_l&=&\frac{\left(R_{1}+R_{2}+\text{i} \omega R_{1} R_{2} C\right) \pi r_{l}^{2}}{1+\text{i} \omega C R_{2}}, \quad \mathrm{for \ physiological \ RCR \ boundary}.
    \label{eq:bc}
\end{subeqnarray}
Here, $c_l$ is the wave velocity at the outlet. It is worth emphasizing that $Z_i$ is normalized by local characteristic impedance, which is $Z_c=\rho c_0$. For the case with non-reflecting boundary condition, the input impedance at low-frequency band is almost four times that of the local characteristic impedance even without any reflection from the outlet. This disparity can be attributed to the constant reflection of the forward flow due to tapering as well the change in wall properties. As the frequency increases, the input impedance decreases and eventually approaches the local characteristic impedance. The phase of the input impedance with non-reflecting boundary also converges to zero as the frequency increases. This indicates that tapering mainly affects the waves with long wavelength, while waves with short wavelength behave as if tapering does not exist. For the case with RCR boundary condition, the input impedance is much higher than the local characteristic impedance at low-frequency band and is nearly four times of the non-reflecting case. The outlet impedance $Z_l$ of RCR boundary is also plotted in figure \ref{fig:Zi-alpha-1}. It is clear that RCR boundary induces a much greater increment in the magnitude of the input impedance than its own magnitude. As the frequency increases, the input impedance of the RCR case does not reduce to zero, but rather oscillates around $Z_c$ as is shown in both the magnitude and the phase plots. Therefore, the inclusion of a physiologically accurate outlet boundary condition is very crucial in the study of the input impedance of arteries.

The effect of $\alpha$ is shown in figure \ref{fig:Zi-all}. For cases with non-reflecting outlet, the behavior of the input impedance at high-frequency band is unaffected by the profile of the wall properties, approaching the local characteristic impedance asymptotically. As is shown in appendix \ref{app:high-freq}, when $\omega \rightarrow \infty$, $\varepsilon \rightarrow \infty$. Substitute $Z_l = \rho c_l$ into equation \ref{eq:Zi_high}, we have $Z_i \approx  \rho c_0$ when $\omega \rightarrow \infty$. Otherwise, when the RCR boundary condition is employed, the input impedance oscillates around the local characteristic impedance at high frequencies. From equation \ref{eq:Zi_high}, it can be shown that as $\omega \rightarrow \infty$, we have
\begin{equation}
    \bigg|\frac{Z_i}{Z_c}\bigg|_{max} = \max \left\{ \frac{R_1\pi r_l^2}{\rho c_l}, \frac{\rho c_l}{R_1\pi r_l^2} \right\}, \quad \text{and} \quad
    \bigg|\frac{Z_i}{Z_c}\bigg|_{min} = \min \left\{ \frac{R_1\pi r_l^2}{\rho c_l}, \frac{\rho c_l}{R_1\pi r_l^2} \right\} ~ \text{at} ~ z=0.
    \label{eq:infin-limit}
\end{equation}
The accuracy of this asymptotic relation is confirmed numerically. It is clear that the oscillation amplitude is determined by the proximal resistance $R_1$ and $\alpha$.

On the other hand, as the frequency decreases, the input impedance converges to the same value regardless of $\alpha$ when RCR boundary is used, while its value decreases as the $\alpha$ value increases when non-reflecting boundary is used. It can be proven (see appendix \ref{app:zero-freq}) that as $\omega \rightarrow 0$
\begin{subeqnarray}
    \frac{Z_i}{Z_c}\bigg| _{z=0} &\approx& \left( \frac{r_0}{r_l} \right)^{\frac{5-\alpha}{2}}, \quad \mathrm{for \ non-reflecting \ boundary}, \\
    \frac{Z_i}{Z_c}\bigg| _{z=0} &\approx& \frac{\pi r_0^2}{\rho c_0} (R_1+R_2), \quad \mathrm{for \ physiological \ RCR \ boundary}.
    \label{eq:zero-limit}
\end{subeqnarray}
Equation \ref{eq:zero-limit}a clearly shows that $Z_i$ is determined by tapering and wall properties jointly, while equation \ref{eq:zero-limit}b shows that RCR boundary condition is the determining factor when it is present. The normalized input impedances predicted by these two equations are also listed in figure \ref{fig:Zi-all}. Compared with values evaluated numerically at $5\times 10^{-3}Hz$, the maximum relative difference is less than 0.1\%, affirming the accuracy of the asymptotic analysis.

The input impedance predicted by the current model (figure \ref{fig:Zi-all}b) are in qualitative agreement with in vivo measurements \citep{nichols1977,murgo1980}. \citet{murgo1980} measured the aortic input impedance in 18 healthy man and noticed the same trend that $|Z_i|$ achieved its maximum at low-frequency, decreased sharply and started to oscillate between $6-8Hz$. Unlike the current study, the oscillation amplitude decreased with increasing frequency due to viscous damping in the arterial wall.

\subsection{Wave reflection}

\begin{figure}
    \centerline{\includegraphics[width=1.0\textwidth]{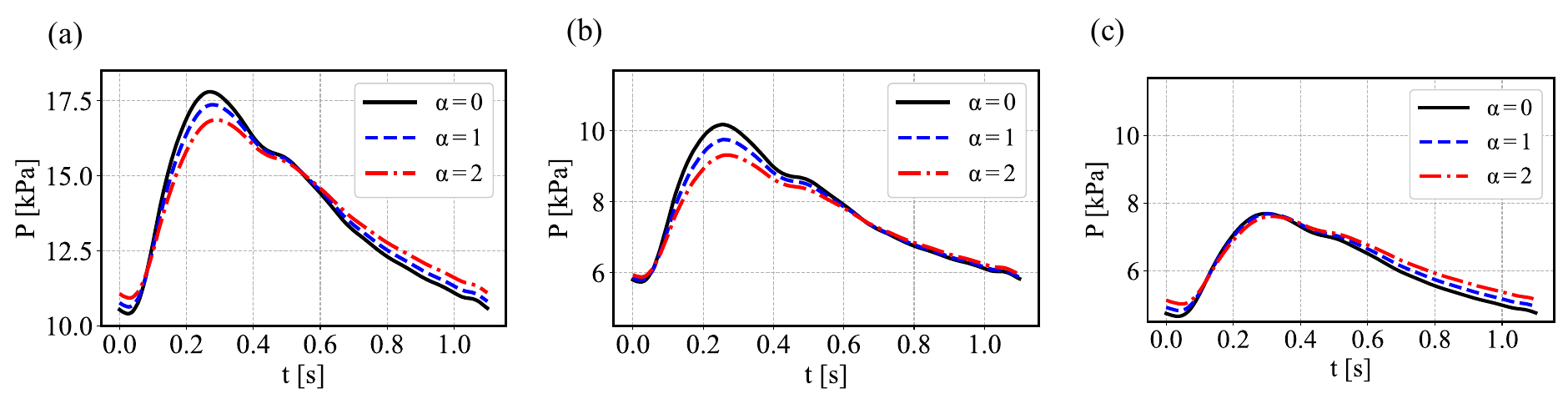}}
    \caption{Pressure waves at $z=l/2$ of the tapered carotid artery model with RCR boundary condition and different $\alpha$.  (a) Total pressure wave; (b) forward wave; (c) backward wave. The $y$-axis of the forward and backward waves is set to the same range to facilitate comparison.}
    \label{fig:wavesplit}
\end{figure}

Tapering, wall property variation and outlet impedance all cause pulse wave reflections. The pressure wave can be separated into forward and backward components using \citep{westerhof1972}
\begin{equation}
    p_{f, b}=\frac{1}{2}(p \pm \rho c v).
    \label{eq:separation}
\end{equation}
Pressure waves at mid-section and its components are compared in figure \ref{fig:wavesplit}. As the $\alpha$ value increases, the artery becomes more compliant, which leads to a decrease in the pulse pressure (difference between the maximum and minimum). This decrease is a combined result of an increase in diastolic pressure (minimum) and a decrease in systolic pressure (maximum). From figures \ref{fig:wavesplit}b and c, it is clear that the forward wave is mostly responsible for the reduction of peak value while the diastolic value is mostly raised by the backward wave. We can also observe a slight temporal shift of the peak pressure value as $\alpha$ changes due to the change in pulse wave velocity.

\begin{figure}
    \centerline{\includegraphics[width=0.9\textwidth]{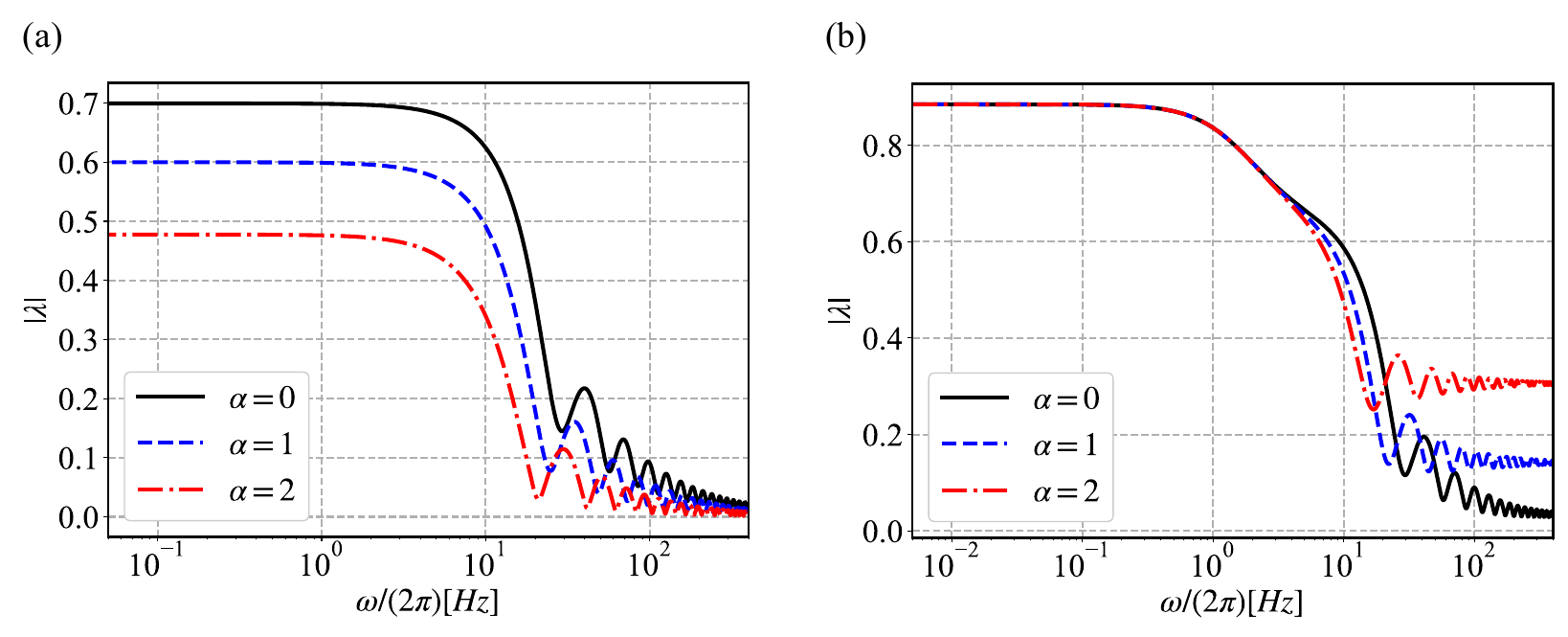}}
    \caption{Reflection coefficient at the inlet of the tapered carotid artery model with different $\alpha$.  (a) Non-reflecting boundary condition is employed; (b) RCR boundary condition is employed.}
    \label{fig:reflection}
\end{figure}
Equation \ref{eq:separation} can also be applied to each frequency component and we have
\begin{align}
    P_{f} &= \frac{1}{2} V_{i n} B r_{0}^{\frac{\alpha}{2}} r^{-\frac{1}{2}} \left(-\frac{I_p(\varepsilon)}{I_v\left(\varepsilon_{0}\right)} \text{i}+\frac{I_v(\varepsilon)}{I_v\left(\varepsilon_{0}\right)}\right), \\
    P_{b} &= \frac{1}{2} V_{i n} B r_{0}^{\frac{\alpha}{2}} r^{-\frac{1}{2}} \left(-\frac{I_p(\varepsilon)}{I_v\left(\varepsilon_{0}\right)} \text{i}-\frac{I_v(\varepsilon)}{I_v\left(\varepsilon_{0}\right)}\right).
\end{align}
The reflection coefficient can be defined as \citep{reymond2009,westerhof2010}
\begin{equation}
    \lambda=\frac{P_{b}}{P_{f}}=\frac{\text{i} I_p/I_v +1}{\text{i} I_p/I_v -1} =\frac{Z_i/Z_c - 1}{Z_i/Z_c + 1}
\end{equation}
Here, $\lambda$ is complex indicating the phase difference between the forward and backward waves. From figure \ref{fig:reflection}, we can see that the behaviors of the reflection coefficient are mostly similar to the input impedance in figure \ref{fig:Zi-all}. One interesting trend is that at high-frequency range with RCR boundary, the reflection coefficient increases with $\alpha$ indicating a growing relative contribution from the backward waves.

\section{Application of the analytic solution} \label{sec:application}

Though developed based on an idealized model, equation \ref{eq:time_domain_solution} can be applied to complex, patient-specific cases. Here we demonstrate the application of the analytic solution to a patient-specific aorta and compare with the results from 3D numerical simulations using CMM.

\subsection{Analytic solution for bifurcation}

\begin{figure}
  \centerline{\includegraphics[width=0.9\textwidth]{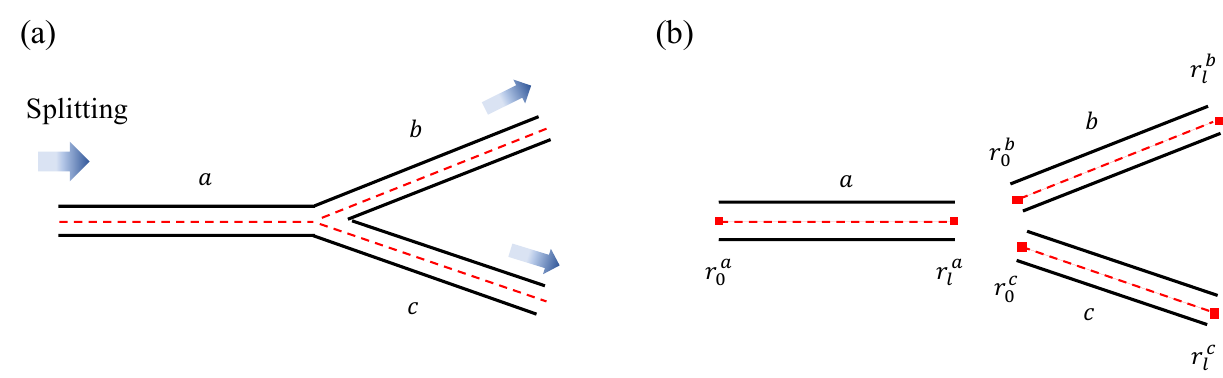}}
  \caption{(a) Schematic of a bifurcation with splitting flow. (b) Decomposition of the bifurcation into three individual vessels. Here, the subscripts $0$ and $l$ represent the inlet and outlet of each vessel, respectively.}
\label{fig:junction}
\end{figure}

The analytic solution presented in this study can be extended to complex models with multiple outlets by decomposing the model into simple blocks that are easier to solve. Similar strategies are adopted in distributed lumped parameter models \citep{mirramezani2022} and 1D models \citep{vandevosse2011}. One of the most common building blocks in an arterial network is the bifurcation. Figure \ref{fig:junction}a shows a typical bifurcation where a parent vessel (labeled $a$) is connected to two daughter vessels (labeled $b$ and $c$). Each vessel in figure \ref{fig:junction}b can be solved with the analytic solution given the proper inlet and outlet boundary conditions, and their solutions are related by the following conditions at the junction \citep{olufsen1999}
\begin{subeqnarray}
    &P_{l}^a=P_{0}^b=P_{0}^c, \\
    &V_{l}^a \pi \left(r_{l}^a\right)^{2}=V_{0}^b \pi \left(r_{0}^b\right)^{2} + V_{0}^c \pi \left(r_{0}^c\right)^{2}.
    \label{eq:junction}
\end{subeqnarray}
Here, the subscripts $0$ and $l$ represent the inlet and outlet of each vessel. For vessel $a$, velocity is prescribed at the inlet and a proper outlet boundary condition, $Z_{l}^a$, is required to obtain its solution. Divide equation \ref{eq:junction}b by equation \ref{eq:junction}a, we get
\begin{equation}
    \frac{\pi \left(r_{l}^a\right)^{2}}{Z_{l}^a}=\frac{\pi \left(r_{0}^b\right)^{2}}{Z_{0}^b}+\frac{\pi \left(r_{0}^c\right)^{2}}{Z_{0}^c}.
    \label{eq:imped_bif}
\end{equation}
The outlet impedance of the vessel $a$ is determined by the input impedances of vessel $b$ and $c$. If vessel $b$ and $c$ are terminal vessels, i.e. they are connected to RCR models, $Z_0^b$ and $Z_0^c$ can be determined explicitly through equation \ref{eq:input_imped}. Then, $Z_l^a$ is readily available through equation \ref{eq:imped_bif}, and the velocity and pressure along vessel $a$ can be obtained. It is worth noting that equation \ref{eq:imped_bif} essentially describes that the input impedance of vessel $b$ and $c$ are connected in parallel to form the outlet impedance of vessel $a$.

For vessel $b$, once the pressure value at its inlet is known, the inlet velocity of vessel $b$ can be obtained from equation \ref{eq:solution}b. Therefore, the solution in this vessel is written as
\begin{align}
    V^b(z)&=\text{i} \frac{P_{l}^a  }{B^b} \left( r_{0}^b \right) ^{\frac{1}{2}} \left( r^{b} \right)^{-\frac{\alpha}{2}} \frac{I_v \left(\varepsilon^b \right)}{I_p \left(\varepsilon^b_0 \right)} \\
    P^b(z)&=P_{l}^a\left(\frac{r_{0}^b}{r^b}\right)^{\frac{1}{2}} \frac{I_p \left(\varepsilon^b \right)}{I_p \left(\varepsilon^b_0 \right)}
\end{align}
The same calculation can be carried out for vessel $c$. This procedure can be expanded to multiple layers of bifurcations as well as junctions with more than two daughter vessels.

Similar to the current study, \citet{flores2016} proposed an analytic solution based on the generalized Darcy’s elastic model in the frequency domain and successfully applied it to model blood flow in complex arterial networks. However, the vessel was assumed to be cylindrical and to have uniform material properties. The tapering and material variation in a large network were modeled in a discrete manner by dividing long vessels into segments. Pressure values at segment ends were treated as unknown variables and were obtained by solving a matrix system constructed from these segments. In the current study, tapering and material changes are built into the analytic solution. The solution process described above is much simpler and can be considered a special case of the matrix-based method when the network only contains splitting junctions.

\subsection{Application to patient-specific aorta}

\begin{figure}
  \centerline{\includegraphics[width=0.8\textwidth]{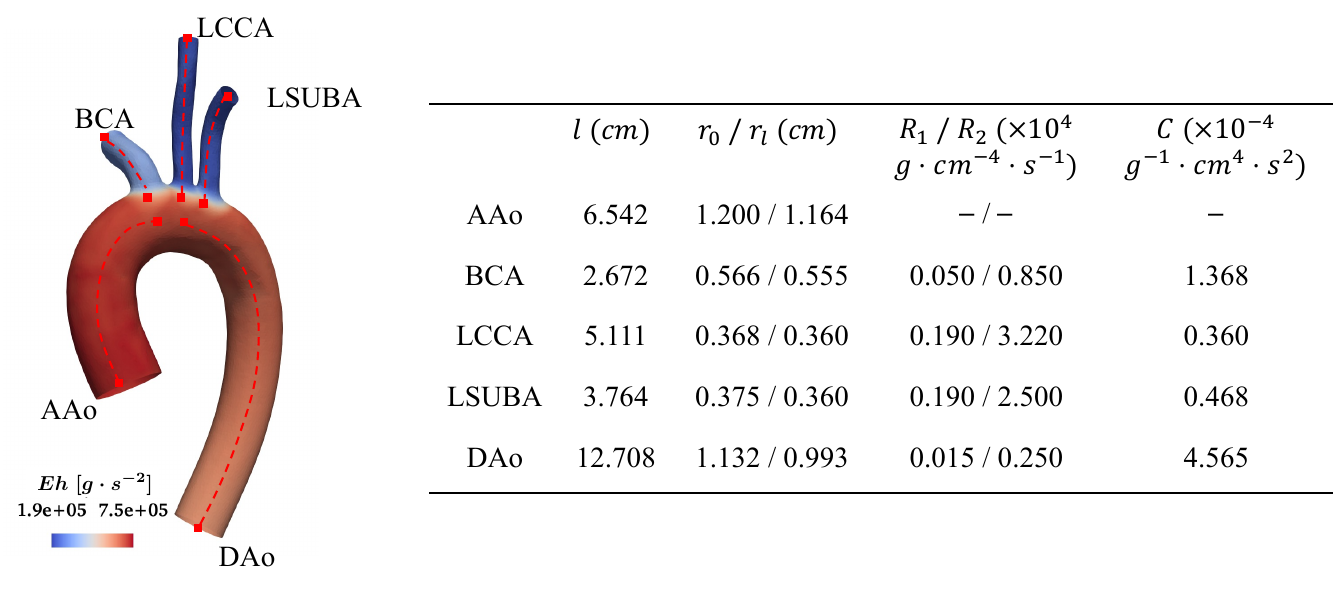}}
  \caption{Patient-specific aorta model and the parameters used in the study. The dashed lines indicate all the sections of arteries used in the analytic model. AAo: Ascending Aorta; BCA: Brachiocephalic Artery; LCCA: Left Common Carotid Artery; LSUBA: Left Subclavian Artery; DAo: Descending Aorta.}
\label{fig:aorta}
\end{figure}

We use a patient-specific aorta model to demonstrate the accuracy and effectiveness of the analytic solution. The model is from an open source dataset \citep{bodyparts3d} and is shown in figure \ref{fig:aorta}. It includes the aorta and three main branches and is broken into individual sections indicated by the dashed lines for analytic modeling. The parameters used in each section are also listed in the figure. The vessel length $l$ is defined as the length of the curved centerline. The variation of the material properties follows the linear relation in figure \ref{fig:Eh}, i.e. $Eh=\beta r$ with $\beta=5.94 \times 10^{5} g \cdot cm^{-1} \cdot s^{-2}$. A pulsatile velocity profile with $T=0.9s$ is prescribed at AAo, and RCR boundary conditions are applied at all of the outlets. In CMM simulations, a grid independence study is carried out and around 0.5 million tetrahedral elements are used to obtained the final results. Based on the Womersley number ($Wo=r \sqrt{2\pi \rho f/\mu} \approx 16$) and the Reynolds number defined with Stokes layer thickness ($Re_{\delta}=\sqrt{2\rho}v_{max}/\sqrt{\mu\omega} \approx 195$), the flow has not transitioned to turbulence under the conditions considered here \citep{merkli1975,hino1976}.

\begin{figure}
    \centerline{\includegraphics[width=0.9\textwidth]{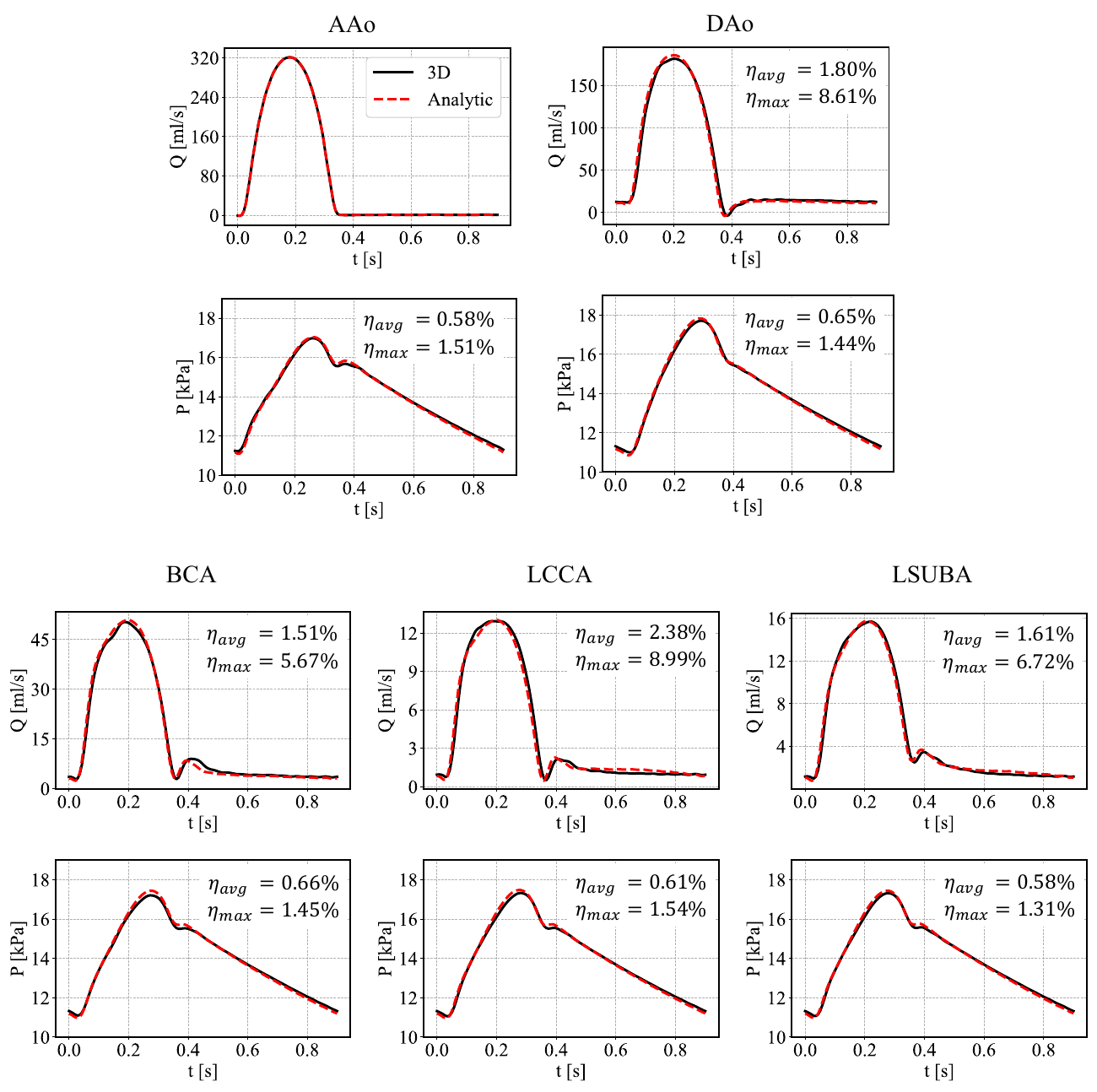}}
    \caption{Comparison between the analytic solutions and 3D simulation results for the patient-specific aorta model.}
  \label{fig:comp_aorta}
\end{figure}

The flow rate and pressure at the inlet and outlets are summarized in figure \ref{fig:comp_aorta}. It can be seen that analytic results are in good agreement with numerical results. The pressure distribution is particularly well-matched, as the maximum relative error is maintained below $2 \%$ and the average relative error remains under $1\%$. The average relative error of the flow rate is less than $2.4 \%$, while the maximum error is higher due to a slight phase difference between these two results. It is worth noting that analytic results can be obtained within 1 second on a desktop equipped with an Intel Core i9-12900K processor, while 3D simulations take approximately 20 minutes per cardiac cycle when run in parallel on 288 Intel Xeon Platinum 9242 cores. Therefore, the analytic solution provides a fast and accurate alternative to 3D simulations in estimating the pressure distribution in this model.


\section{Conclusion} \label{sec:conclusion}

In this study, we derive an analytic solution for pulse wave propagation in an arterial model using frequency domain analysis. In addition to tapering, this model also includes variable wall properties that follow the profile $Eh=\beta r^\alpha$ and a physiological RCR outlet boundary condition that models the resistance and compliance of the downstream vascular network. This analytic solution is successfully validated against 1D and 3D numerical simulations. Then, it is used to theoretically analyze the wave propagation characteristics in an idealized model. It is confirmed that tapering and variable wall properties can create constant reflections along the path. Our study also demonstrates that wall properties and RCR boundary condition have a significant impact on the wave propagation, and their influences are particularly prominent at the low-frequency range. Even though it is observed in figure \ref{fig:Zi-all} that high-frequency components are also affected by these factors, it is essential to approach these findings with caution, as viscous effect is not considered in the current model. Furthermore, it is worth noting that these high-frequency components may not hold significant physiological relevance, given the intrinsic frequency of the cardiac cycle.

Moreover, the analytic solution is applied to rapidly and accurately estimate the pressure distribution in a patient-specific aorta by splitting the model into individual sections and applying the analytic solution to each section. Compared to numerical methods, the analytic solution can be a computationally economical alternative for modeling pulse wave propagation. It also enables theoretical analysis to quantify the influence of different model parameters, such as boundary conditions and material properties, thus allowing for quick tuning of these parameters, which can then be used in 1D and 3D numerical simulations. Additionally, this method is potentially useful for clinical applications such as the estimation of the central pressure from peripheral pressure measurements. Building upon the study by \citet{flores2021}, it is possible to model the pressure wave propagation from the aorta to the brachial artery using the current analytic solution and derive a transfer function between central and brachial pressure.

There are several limitations in the current study. First, the omission of the nonlinear term (except for the steady component) and blood viscosity in the momentum equation leads to the under-estimation of pressure values. \citet{reymond2009} demonstrated that both effects contributed about single-digit percentages to the predicted pressure value. Given that the errors we observe are of the same order of magnitude, including these effects can potentially improve our results. This is especially important in predicting pulse wave propagation within a long artery network with multiple layers of bifurcations, in which case avoiding error accumulation is of greater importance. Second, the blood vessel is more accurately modeled as a viscoelastic material. Experimental evidences have shown that there is hysteresis between pressure and lumen area \citep{valdez-jasso2009} and the viscoelasticity causes attenuation of pulse waves as they travel downstream \citep{bessems2008}. Last but not least, the current model cannot be applied directly to diseased arteries such as those with aneurism or stenosis, but can potentially be expanded to model these anomalies \citep{papadakis2019}.

\paragraph{Acknowledgements.} The authors thank Dr. Baofang Song for valuable discussions.
\paragraph{Funding.} This work is supported by the National Key Research and Development Program of China (grant nos 2021YFA1000200, 2021YFA1000201) and the National Natural Science Foundation of China (grant no. 12272009). C.Z. also received financial support from Fundamental Research Funds for the Central Universities, Peking University (grant nos 7100604109, 7100604343) and Young Elite Scientists Sponsorship Program by BAST (grant no. BYESS2023025).
\paragraph{Declaration of interests.} The authors report no conflict of interest.
\paragraph{Author ORCIDs.} C. Zhu, https://orcid.org/0000-0002-1099-8893

\appendix

\section{High-frequency limit}
\label{app:high-freq}

When $\omega \rightarrow \infty$, we have $\varepsilon \rightarrow \infty$. For Bessel functions at $\varepsilon \rightarrow \infty$, we have \citep{abramowitz1948}
\begin{equation}
    J_{\nu}(\varepsilon) \approx \sqrt{\frac{2}{\pi \varepsilon}} \cos \left(\varepsilon-\frac{v}{2} \pi-\frac{\pi}{4}\right), \quad Y_{\nu}(\varepsilon) \approx  \sqrt{\frac{2}{\pi \varepsilon}} \sin \left(\varepsilon-\frac{v}{2} \pi-\frac{\pi}{4}\right).
    \label{eq:bessel_infty}
\end{equation}
From equation \ref{eq:Zl}, the impedance at the outlet can be simplified to a real value
\begin{equation}
    Z_l \approx R_1\pi r_l^2.
\end{equation}
Define
\begin{equation}
    R_l = \frac{Z_lr_l^{\frac{1-\alpha}{2}}}{B} = \frac{Z_l}{\rho c_l}
    \label{eq:Rl}
\end{equation}
and substitute the above equations into equation \ref{eq:F}, and we obtain
\begin{equation}
    F \approx - \frac{ \text{i} R_l \sin \left(\varepsilon_l-\frac{v}{2} \pi-\frac{\pi}{4}\right) - \cos \left(\varepsilon_l-\frac{v}{2} \pi-\frac{\pi}{4}\right) } { \text{i} R_l \cos \left(\varepsilon_l-\frac{v}{2} \pi-\frac{\pi}{4}\right) + \sin \left(\varepsilon_l-\frac{v}{2} \pi-\frac{\pi}{4}\right) }
\end{equation}

The normalized input impedance as $\varepsilon \rightarrow \infty$ can be simplified to
\begin{align}
    \frac{Z_i}{Z_c}\bigg| _{z=0} &= -\text{i}\frac{I_p}{I_v}\bigg| _{z=0} \nonumber \\
    &\approx \left[ \frac{R_l}{R_l^2 \sin ^2 \theta + \cos ^2 \theta } + \text{i}\frac{ \left( R_l^2-1 \right)  \sin \theta \cos \theta }{R_l^2 \sin ^2 \theta + \cos ^2 \theta } \right] \bigg| _{z=0}.
    \label{eq:Zi_high}
\end{align}
Here, $\theta=\varepsilon_l-\varepsilon$. If $Z_l = \rho c_l$, we have $R_l=1$ and $Z_i\approx Z_c$ at the inlet. If $R_l\ne 1$, it can be shown the period of $|Z_i/Z_c|$ is $\theta = \pi$ and the maximum and minimum are $\max \left\{ R_l , 1/R_l \right\}$ and $\min \left\{ R_l , 1/R_l \right\}$, respectively.

\section{Low-frequency limit}
\label{app:zero-freq}

In the limit $\omega \rightarrow 0$, we have $\varepsilon \rightarrow 0$ and the following relation for Bessel function \citep{abramowitz1948}
\begin{equation}
    J_{\nu}(\varepsilon) \approx \frac{1}{\Gamma(v+1)}\left(\frac{\varepsilon}{2}\right)^{v}, \quad Y_{\nu}(\varepsilon) \approx -\frac{\Gamma(v)}{\pi}\left(\frac{\varepsilon}{2}\right)^{-v}.
    \label{eq:bessel_zero}
\end{equation}
Moreover, from equation \ref{eq:Zl}, the impedance at the outlet can be simplified to a real value
\begin{equation}
    Z_l \approx (R_1+R_2)\pi r_l^2.
\end{equation}
Substitute the above relationships along with equation \ref{eq:Rl} into equation \ref{eq:F}, and we obtain
\begin{equation}
    F = \underbrace{- \frac{R_l^2 J_{\nu} Y_{\nu} + J_{\nu-1}Y_{\nu-1}}{R_l^2 J_{\nu}^2 + J_{\nu-1}^2}}_{F_r} + \text{i} \underbrace{\frac{R_l ( J_{\nu-1} Y_{\nu} - J_{\nu}Y_{\nu-1} )}{R_l^2 J_{\nu}^2 + J_{\nu-1}^2}}_{F_i}.
\end{equation}
Note that the Bessel functions are evaluated at $\varepsilon_l$ here. From the asymptotic relation in equation \ref{eq:bessel_zero} and the parameters in table \ref{tab:para_carotid}, it can be shown that the following equations hold in this study
\begin{equation}
    |F_r| \ll |F_i|, \quad  |J_{\nu}Y_{\nu-1}| \ll |J_{\nu-1} Y_{\nu}|. \nonumber
\end{equation}
Hence, in the limit of $\varepsilon \rightarrow 0$, we have
\begin{equation}
    F \approx \text{i} F_i,
\end{equation}
and
\begin{equation}
    F_i \approx \frac{R_l ( J_{\nu-1} Y_{\nu} )}{R_l^2 J_{\nu}^2 + J_{\nu-1}^2} \approx - \frac{R_l\nu^2\Gamma(\nu)^2}{\pi} \left[R_l^2  \left(  \frac{\varepsilon_l}{2} \right)^{2\nu+1} + \nu^2 \left( \frac{\varepsilon_l}{2} \right)^{2\nu-1} \right]^{-1}.
\end{equation}
The recursive relation $\Gamma(v+1)=v\Gamma(v)$ is used in the above derivation.

The normalized input impedance as $\varepsilon \rightarrow 0$ can be simplified to
\begin{equation}
    \frac{Z_i}{Z_c}\bigg| _{z=0} = -\text{i}\frac{I_p}{I_v}\bigg| _{z=0} \approx  F_i \frac{J_{\nu-1}}{Y_{\nu}} \bigg| _{z=0}.
\end{equation}
Further simplify the above equation, we obtain
\begin{equation}
    \frac{Z_i}{Z_c}\bigg| _{z=0} \approx R_l \left( \frac{\varepsilon_0}{\varepsilon_l} \right)^{2\nu-1}.
    \label{eq:Zi_zero}
\end{equation}
It is verified that the asymptotic equation \ref{eq:Zi_zero} is valid for both non-reflecting boundary and physiological RCR boundary. In both cases, as $\varepsilon \rightarrow 0$, we have
\begin{subeqnarray}
    \frac{Z_i}{Z_c}\bigg| _{z=0} &\approx& \left( \frac{r_0}{r_l} \right)^{\frac{5-\alpha}{2}}, \quad \mathrm{for \ non-reflecting \ boundary}, \\
    \frac{Z_i}{Z_c}\bigg| _{z=0} &\approx& \frac{\pi r_0^2}{\rho c_0} (R_1+R_2), \quad \mathrm{for \ physiological \ RCR \ boundary}.
    \label{eq:Zi_zero2}
\end{subeqnarray}
where $c_0$ is the pulse wave velocity at the inlet. Equation \ref{eq:Zi_zero2}b is consistent with the steady state solution (equation \ref{eq:steady_solution}), neglecting the nonlinear effect.

\bibliographystyle{unsrtnat}

\end{document}